\documentclass{elsart}
\usepackage{graphicx,harvard,amssymb}

\makeatletter
\def\elsartstyle{%
	\def\normalsize{\@setfontsize\normalsize\@xiipt{14.5}}
	\def\small{\@setfontsize\small\@xipt{13.6}}
	\let\footnotesize=\small
	\def\large{\@setfontsize\large\@xivpt{18}}
	\def\Large{\@setfontsize\Large\@xviipt{22}}
	\skip\@mpfootins = 18\p@ \@plus 2\p@
	\normalsize
}
\makeatother

\def\url#1{{\ttfamily\def\/{/\discretionary{}{}{}}#1}}

\begin{document}
\begin{frontmatter} 
\title{Tracking through equality}

\author{Giandomenico Sassi$^{1}$, Silvio A. Bonometto$^{1,2}$}
\address{1 -- Physics Department G. Occhialini, Universit\`a degli Studi di
Milano--Bicocca, Piazza della Scienza 3, I20126 Milano (Italy)}
\address{2 -- I.N.F.N., Sez.~Milano--Bicocca,
Piazza della Scienza 3, I20126 Milano (Italy)}

\thanks[email]{E-mails: sassi@mib.infn.it, bonometto@mib.infn.it}

\begin{abstract}
We give a tracker solution for the quintessence scalar field for
Ratra--Peebles or SUGRA potentials, holding before, during and after
the equality epoch ($\rho_m=\rho_r$) and nicely fitting the
numerical behavior.
\end{abstract}


\end{frontmatter}

\maketitle
\section{Introduction}
Dark Energy (DE) is one of the main puzzles of cosmology. It could be
a scalar field $\phi$, self--interacting through a potential $V(\phi)$
(Wetterich 1988, Ratra \& Peebles 1988), so that
\begin{equation}
\rho_{DE}=\rho_{k,DE}+\rho_{p,DE} \equiv {{\dot{\phi }}^{2}/2a^{2}}+V(\phi ),~
~~~~
p_{DE}=\rho_{k,DE}-\rho_{p,DE} = w \rho_{DE}~,
\end{equation}
provided that $\rho_{k,DE}/V\ll 1/2$, so that $-1/3 \gg w >-1$. Here
\begin{equation}
ds^{2} = a^{2}(\tau) (-d\tau^{2}+ dx_i dx^i ) ~,~~~~~ (i=1,..,3)
\end{equation}
is the metrics and dots indicate differentiation with respect to $\tau
$ (conformal time). This kind of DE is dubbed {\it dynamical} DE (dDE)
or {\it quintessence}.

The most significant potentials are those allowing tracker solutions,
so limiting the impact of initial conditions. Among them, we consider
here the SUGRA (Brax \& Martin 1999, 2001; Brax, Martin \& Riazuelo
2000) and Ratra \& Peebles (1988, 2003) potentials
\begin{equation}
\label{ptntl}
SUGRA ~:~~~
V(\phi) = (\Lambda^{4+\alpha}/\phi^\alpha) \exp(4 \pi \phi^2/m_p^2)
\end{equation}
\begin{equation}
RP ~:~~~
V(\phi) = \Lambda^{4+\alpha}/\phi^\alpha
\end{equation}
($m_p=G^{-1/2}$: Planck mass); in the recent times, they yield $|w|$
close to unity and fastly variable or $|w|$ farther from unity and
slowly variable, respectively. In a flat model, once the DE density
parameters $\Omega_{DE}$ is assigned, either $\alpha$ or the energy
$\Lambda$ can still be freely chosen.

SUGRA and RP behave differently when $\phi$ approaches $m_p$. Here,
however, we shall not deal with late epochs and our expressions,
worked out for RP, suit SUGRA as well.

Tracker solutions are then usually found by considering the equation
of motion
\begin{equation}
\ddot \phi + 2(\dot a/a) \dot \phi = \alpha a^2 \Lambda^{4+\alpha}/
\phi^{1+\alpha}
\label{phieq}
\end{equation}
and seeking solutions of the form
\begin{equation}
\phi = \phi_i (\tau/\tau_i)^\beta
\end{equation}
($\tau_i$ is a reference time and $\phi_i$ is the field value at that
time). They are fixed once the time dependence of the scale factor
$a(\tau)$ is set. In turn, we can however write
\begin{equation}
\dot a/a = u(\tau)/\tau~,
\end{equation}
so that $u=1$ ($u=2$) in the radiation (matter) dominated eras and is
actually $\tau$ dependent around matter--radiation equality.

In this note we give a single expression for the tracker solution
holding before, during and after equality, occurring at $\tau_e$. This
expression improves the tracker solution expression, known for $\tau
\ll \tau_{e}$, when $\tau$ approaches $\tau_e\, $. It also neatly
improves the tracker solution known for $\tau \gg \tau_{e}$, for a
large range of redshifts.  Furthermore, it fits the behavior of the
$\phi$ field across equality, with quite a small discrepancy from the
numerical solution, keeping mostly well below $1\, \%\, $.

\section{Tracker solutions through equality}
Let be $R = a/a_{e}$ and $\theta = \tau/\tau_{e}$ ($a_e$, $\tau_e$:
equality scale factor, time); it is easy to verify that the expression
\begin{equation}
R = c^2 \theta^2 + 2\, c\, \theta~~~~~~~~~~~~~~{\rm with}~~c=\sqrt{2}-1~,
\label{rtheta}
\end{equation}
gives the scale factor behavior before, during and after equality.
This follows the integration of the Friemann equation 
\begin{equation}
\dot R = H_e a_e (R+1)^{1/2}
\label{frw}
\end{equation}
($3H_e = 8\pi G \rho_{e,m} = 8\pi G \rho_{e,r}$; $\rho_{e,m}=
\rho_{e,r}$ are matter and radiation energy densities at equality) and
noticing that $H_e a_e = 2c/\tau_e\, $.  It is then also easy to see
that
\begin{equation}
\label{utheta}
u(\theta) = {1 + c\theta \over 1+c\theta/2}
\end{equation}
so that, clearly, $u \to 1$ ($u \to 2$) for $\theta \to 0$ ($\theta
\to \infty$).

Let us now rewrite eq.~(\ref{phieq}) in the form
\begin{equation}
\label {yeq}
Y'' + 2{u \over \theta} Y' = \alpha R^2 Y^{-(1+\alpha)}~.
\end{equation}
Here $Y = \phi/\sigma$ with $\sigma^{2 + \alpha} = (a_e \tau_e)^2
\Lambda^{4+\alpha}$ and $'$ indicates differentiation in respect to
$\theta$. Let us then seek solutions of eq.~(\ref{yeq}) of the form
\begin{equation}
Y = Y_e \, (R\, \theta)^b ~~~{i.e.}~~~\phi = \sigma Y_e
(a\tau/a_e\tau_e)^b~.
\label{newtra}
\end{equation}
Clearly
\begin{equation}
\label{deriva}
Y' = {b \over \theta}\, (1+u) Y
~,~~~~
Y'' = {1 \over \theta^2}
\left[\, b^2 (u+1)^2 + bu'\theta - b(u+1) \right] Y
\end{equation}
so that eq.~(\ref{yeq}) yields
\begin{equation}
Y_e^{2+\alpha} (R\, \theta)^{b(2+\alpha)} = \alpha (R\, \theta)^2 G(\theta)
\end{equation}
Here
\begin{equation}
G^{-1} (\theta) =  b^2 (u+1)^2 + b(u+1)(2u-1) + bu'\theta 
\label{gm1}
\end{equation}
depends on $\theta$, however not so strongly as is meant by a power
law. Let then be
\begin{equation}
b = 2/(2+\alpha)~,~~~
Y_e = \alpha G~,
\label{bye}
\end{equation}
although this means that also $Y_e$ depends on $\theta$ and,
therefore, the expressions (\ref{deriva}) should contain further
terms. The $\theta$ dependence of $G$ arises from the $\theta$
dependence of $u$; should we neglect the former one, the last term in
the expression (\ref{gm1}) should be consistently omitted.

In the next section we shall compare the behavior of $\phi$, as
obtained from the expressions (\ref{newtra}), (\ref{gm1}),
(\ref{bye}), denominated GT, with a numerical solution of
eq.~(\ref{phieq}).

\section{High and low $\theta$ limits}
Let us first consider the solution (\ref{newtra}) in the limit $\theta
\ll 1$. Then $u \simeq 1$, $G \simeq 1/[(2b)^2+(2b)]$ and, according
to eq.~(\ref{rtheta}), $R\simeq 2c\theta$.  Accordingly, we obtain
\begin{equation}
Y_r = \left[ 4\, c^{2} \alpha \over \beta_r^2+\beta_r\right]^{1 \over
2+\alpha} \theta^{\beta_r} ~~,~~~~~ {\rm with}~~ \beta_r = 4/(2+\alpha)
\label{radtra}
\end{equation}
so that, as expected, we recover the tracker solution for the
radiation dominated regime (herebelow RT).
\begin{figure}
\begin{center}
\vskip -3.5truecm
\includegraphics*[width=12cm]{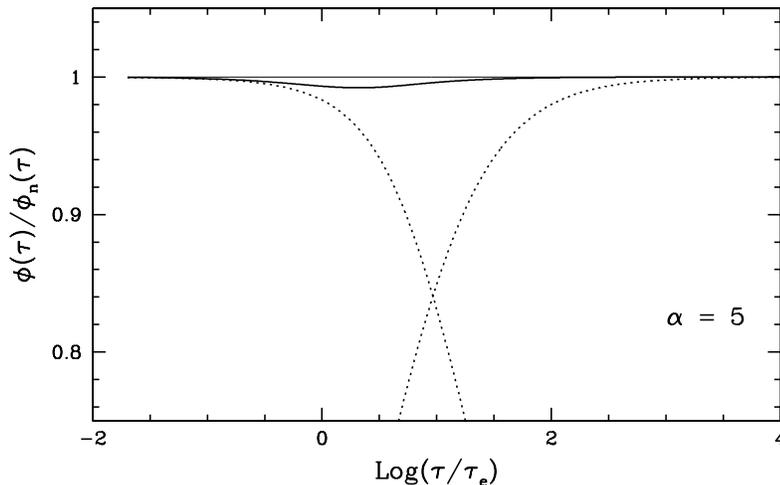}
\end{center}
\vskip -3.2truecm
\caption{Comparison between a the tracker solution $\phi(\tau)$
obtained in this work (GT) and the numerical integral $\phi_n(\tau)$
of the $\phi$--field equation.  Dotted lines show RT and MT solutions
vs.~the numerical solution. The GT is consistent with the numerical
integral, apart a narrow interval, where its discepancy from the
numerical integral marginally approaches $1\, \%\, $.  RT is a fair
approximation until $\tau_{e}$. MT performs much worse.  }
\label{track}
\end{figure}

In quite the same way, for $\theta \gg 1$, we have $u \simeq 2$, $G
\simeq 1/[(3b)^2+3(3b)]$ and, according to eq.~(\ref{rtheta}),
$R\simeq c^2\theta^2$.  Accordingly, we obtain
\begin{equation}
Y_m = \left[ c^4 \alpha \over \beta_m^2+3\, \beta_m \right]^{1 \over
2+\alpha} \theta^{\beta_m} ~~,~~~~~ {\rm with}~~ \beta_m = 6/(2+\alpha)
\label{mattra}
\end{equation}
and, again, we recover the tracker solution for a matter dominated
regime (herebelow MT).

\section{Comparison with numerical integrals}
The tracker solution (\ref{radtra}) can be used to set the initial
conditions to the eq.~(\ref{yeq}), performing then its numerical
integration. In doing so we can use the expression (\ref{rtheta}) and
(\ref{utheta}) so that the system to integrate amounts to 2 first
order equations. We shall indicate the numerical integral by
$Y_n(\theta)\, $. In Figure \ref{track} we plot the ratios $Y/Y_n$,
$Y_r/Y_n$, $Y_m/Y_n\, $.

The symmetry between RT and MT solution is only
apparent. Dotted curves are symmetric in respect to $\tau \simeq 10\,
\tau_{e}$, and this means that at such redshift, inside the matter
dominated era, the RT solution still performs as well as the MT
solution. Using the expressions (\ref{radtra}) and (\ref{mattra})
it is easy to see that the two solutions intersect for
\begin{equation}
\theta^2 = {18 \over c^2} {\beta_r^2/2 + \beta_r 
\over \beta_r^2 + \beta_r}
\end{equation}
{\it i.e.}, for $\theta = \tau/\tau_e \sim \sqrt{18}/(\sqrt{2}-1) \sim
10$ and this confirms what is shown by the plot.
\begin{figure}
\begin{center}
\vskip -3.5truecm
\includegraphics*[width=12cm]{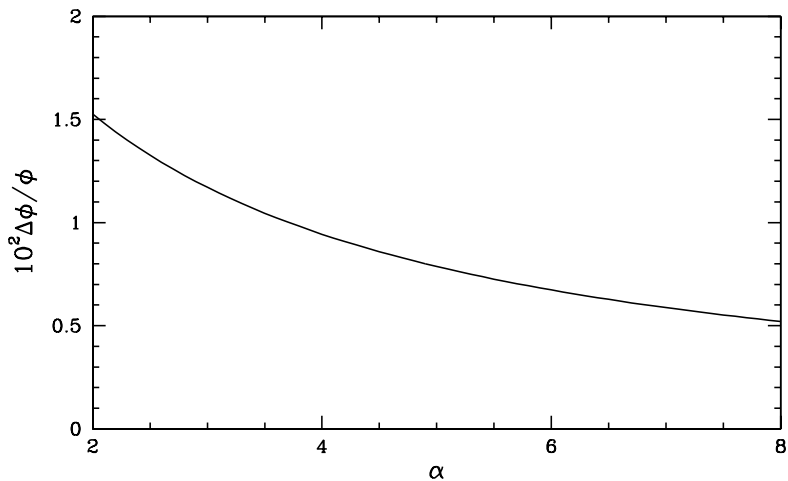}
\end{center}
\vskip -3.2truecm
\caption{Maximum discrepancy between GT and numerical solutions. }
\label{dphimax}
\end{figure}

Altogether, this means that a fair tracker solution, since $\sim
\tau_e$, can only be given by using the GT expression, holding until
DE itself begins to play a significant role as a source of the cosmic
expansion.  The maximum discrepancy $\Delta \phi/\phi$ between GT and
numerical solution is $\sim 1\, \%\, $, slightly after $\tau_e$, for
any reasonable $\alpha$ value. Its $\alpha $ dependence is plotted in
Figure \ref{dphimax}.

\section{Conclusions}
Tracker solutions play an important role in the analysis of dDE
cosmologies. A good tracker solution for the radiation dominated
regime can be easily given. A tracker solution for the matter
dominated regime also exists, but its usual analytical expression is
rather far from a fair numerical behavior.

Here we gave a fair solution performing as well as the usual one in
the radiation dominated era, fitting the numerical solution also about
the equality (the residual discrepancy approaches $\sim 1\, \% \, $
just in a narrow interval) and fitting the numerical solution all
through the matter dominated era. It reads
\begin{equation}
\phi = \phi_e \left( {\tau \over \tau_e} {a \over a_e} \right)^b
\end{equation}
with
\begin{equation}
b={2 \over 2+\alpha}~,~~
\phi_e = {\alpha \Lambda (a_e \tau_e \Lambda)^b
\over b^2 (u+1)^2 - b(u+1)(2u-1) }~.
\end{equation}
It was found by using an exact analytical expression for $a(t) $
through the equality period, also to work out the $\tau$ dependence of
$u = \tau \dot a/a\, $.

When initial conditions to any numerical problem are to be set, one
usually needs to go back to the radiation dominated era, so to rely on
a fair tracking. As an example of problems where initial conditions in
the matter dominated era are useful, let us remind the study of the
evolution of a spherical fluctuation, the prediction of cluster mass
function and its redshift evolution (Mainini, Macci\`o \& Bonometto
2003, Mainini et al. 2003, Mainini 2005) using a Press \& Schechter
(1974) or similar (Sheth \& Tormen 1999, 2002, Jenkins et al, 2001)
approach, and performing N--body simulations (see, e.g., Klypin et
al. 2003, Macci\`o et al. 2004, Solevi et al. 2006).

Using the GT expression, given here, this can be safely avoided, and
initial conditions can be given at any time, until DE itself begins to
be important for the overall cosmic expansion.

\vskip .3truecm

{\bf Acknowledgments}

Loris Colombo and Roberto Mainini are gratefully thanked for
discussions.

{}
\end{document}